# Selective emitters design and optimization for thermophotovoltaic applications

E. Nefzaoui,[*] J. Drevillon, and K. Joulain

*Institut Pprime, CNRS-Université de Poitiers-ENSMA, Département Fluides,*
*Thermique, Combustion, ENSIP-Bâtiment de mécanique, 40,*
*avenue du Recteur Pineau, F 86022 Poitiers, Cedex, France*

(Dated: February 15, 2012)

Among several solutions to exploit solar energy, thermophotovoltaics (TPV) have been popularized and have known great breakthroughs during the past two decades. Yet, existing systems still have low efficiencies since the wavelength range of optimal photovoltaic (PV) conversion is very small compared to the emitter spectral range. Selective emitters are a very promising solution to this problem. We developed numerical tools to design and optimize such emitters. Some of the resulting structures composed of two or four layers of metals and semiconductors are presented in this paper. We also show that the usual PV devices efficiency limits (30% for crystalline silicon under solar radiation, according to Shockley-Queisser model) can be easily overcome thanks to these structures.

---------

[*]elyes.nefzaoui@univ-poitiers.fr; Author to whom correspondence should be addressed.

# I. INTRODUCTION

A TPV device aims to convert thermal energy emitted by a body heated at a temperature $T_e$ into electricity in the same way PV devices convert sun radiation. However, a fundamental difference exists between both systems: in TPV, the emitter temperature is much lower than the sun surface temperature. This implies that the largest part of radiation is emitted in the infrared range of the electromagnetic spectrum instead of the visible range. Low band gap materials (InGaAs, GaSb, etc.), different from those used in PVs (Si), are then needed.

In both cases, efficiency theoretical limits (of PV devices, and TPV devices all the more) have been well known for a few decades[1] and the efficiency of existing devices is still low due to different limiting phenomena involving both emitter and receiver radiative properties. To overcome these limitations, several solutions have already been proposed[2]: the use of complex filters and reflectors systems to recover the non converted radiation, multijunction cells that combine different photosensitive materials adapted to the different regions of the electromagnetic spectrum, etc. A promising solution developed in the last years, and made possible by the development of nanofabrication techniques, focuses on the emitter instead of the receiver (photosensitive material) or the transmitter(filter). It consists on using selective emitters (SEs) adapted to each converter.

A SE is a body that emits all its radiation in a narrow wavelength range well-matched to the receiver band gap wavelength. This idea has become reasonable since it was shown, at the beginning of the last decade, that a thermal radiation very different from that of a blackbody (BB) could be obtained and that its spectral and directional control was possible[3]. Several authors focused on this kind of emitters and proposed genuine solutions based on one dimensional (1D) photonic crystals[4], single-defect photonic crystals[5] or random multilayer structures[6]. Others proposed solutions based on 2D and 3D nanostructured materials [7]. However all these structures are still too complex to consider an industrial process since they are composed of a large number of layers or necessitate a complex surface nanostructuration. Simpler structures based on resonant cavities have also been demonstrated [8, 9].

For this reason, we focused our efforts on simplifying existing structures or designing simpler ones based on 1D multilayer structures. Numerical tools have been developed for this purpose. These tools implement some stochastic optimization methods such as genetic algorithms[6] and particle swarms [10]. First, a simplified calculation of a TPV system efficiency shows that these emitters allow to overcome the usual efficiency limits of single junction cells [1]. Then, some of the obtained multilayer structures, based on dielectrics, metals and semi-conductors are presented in this paper. These structures exhibit quasi-monochromatic thermal emission in near and mid infrared range and are sufficiently simple to consider industrial manufacturing. Besides, they can be easily tuned in order to adapt the emission peak wavelength to different PV cells (InGaAs, GaSb, . . . ), control the emission peak width and obtain different output powers.

# II. A TPV DEVICE EFFICIENCY

## A. A simplified model

In this paragraph, a simplified model for the calculation of a TPV device efficiency is presented. This model was first proposed by Shockley and Queisser [1]. Their notations are adopted here and their model will be considered in the following paragraphs to compare systems with different emitters.

The efficiency $\eta$ is given by the ratio of electric power $P_e$ by incident radiative power $P_r$.

$$\eta = \frac{P_e}{P_r} \qquad (1)$$

The incident radiative power per unit area is given by

$$P_r = A f_\omega \int_0^{2\pi} \int_0^\infty \epsilon_{\lambda,\theta} \ I_{\lambda,\theta}(T_e) \ d\lambda d\Omega \qquad (2)$$

where $A$ is the converter (cell) area, $f_\omega$ is the view factor between the emitter and the cell, $\epsilon_{\lambda,\theta}$ is the emissivity of the emitter and $I_{\lambda,\theta}(T_e)$ the radiation intensity of a BB at a temperature $T_e$. $\lambda$, $\theta$ and $\Omega$ figure the wavelength, the angle of incidence and the solid angle respectively.

The electric power is given by

$$P_e = J \cdot V \qquad (3)$$

where $J$ is the electric current and $V$ is the voltage between the terminals of the $p$-$n$ junction. In order to simplify the electric current calculation, a few assumptions are made:

- All incident photons are absorbed.

- Each photon with an energy higher than the band gap energy $E_g = \frac{hc}{\lambda_g}$ (where $\lambda_g$ is the band gap wavelength) creates one electron-hole pair (quantum efficiency of 1).

- Only radiative recombinations are taken into account (nor Auger neither Shockley-Read-Hall recombinations).

- A view factor equal to that of the sun disk viewed from earth ($f_\omega = 6.85 \times 10^{-5}$sr) is considered.

According to the previous assumptions, we just need to determine the number of incident photons with energy higher than $E_g$ to obtain the number of photo-generated electrons and consequently the electric intensity and power.
The incident flux density of these photons is given by:

$$Q_i = \int_0^{2\pi} \int_0^{\lambda_g} \frac{\epsilon_{\lambda,\theta} \ I_{\lambda,\theta}(T_e)}{(hc/\lambda)} \ d\lambda d\theta \qquad (4)$$

The rate of radiative recombination in the cell under radiation is given by

$$R_c(V) = 2 \cdot A \cdot Qc \cdot exp(\frac{V}{V_c}) \qquad (5)$$

where

$$Q_c = \int_0^{2\pi} \int_0^{\lambda_g} \frac{\epsilon_{\lambda,\theta,c} \ I_{\lambda,\theta}(T_c)}{(hc/\lambda)} \ d\lambda d\theta \qquad (6)$$

and $V$ is the difference in quasi Fermi levels of electrons and holes which is equal to the voltage between the terminals of the $p$-$n$ junction, $V_c$ stands for $k_b T_c/q$ where $k_b$ is Boltzmann constant, $T_c$ is the cell temperature and $q$ is the elementary electric charge and $\epsilon_{\lambda,\theta,c}$ is the cell emissivity.
The current-voltage relationship of the cell is given by

$$J = J_{sh} + J_0(1 - exp(\frac{V}{V_c})) \qquad (7)$$

where $J_{sh}$ is the short circuit current

$$J_{sh} = qA(Q_i - 2Q_c) \qquad (8)$$

and $J_0$ is the reverse saturation current

$$J_0 = qR_c exp(-\frac{V}{V_c}) \qquad (9)$$

### B. Blackbody Vs selective emitters

The previous model is applied to a PV system using direct solar radiation, a TPV system using filtered solar radiation (Fig. 1a) and a SE (Fig. 1b) respectively. We compare the efficiency and the output power of the three systems. Two parameters are considered: the converter band gap energy and the SE emission peak width.

#### 1. Effect of the converter band gap

We calculate the efficiency $\eta$ defined in equation 1. The converter band gap energy $E_g$ is defined by its equivalent wavelength $\lambda_g$. $\lambda_g$ appears in equations 4 and 6. Its values mainly affect the values of absorbed incident radiation (converted energy) and radiative recombination. Solar radiation is modeled by a BB at $T_s = 5800$ K. The BB emissivity is $\epsilon_{BB} = 1$ for all wavelengths. The SE emits in a narrow range below $\lambda_{peak} = 1.0$ $\mu$m at $T_e = 5800$ K. Its emissivity is modeled by a rectangular function between 0.8 $\mu$m and 1.0 $\mu$m as follows:

$$\begin{cases} \epsilon_{SE}(\lambda) = 1 \text{ for } \lambda \in [0.8, 1]\mu m \\ \epsilon_{SE}(\lambda) = 0 \text{ otherwise} \end{cases} \qquad (10)$$

The PV cell is considered at $T_c = 300$ K. The efficiency of the TPV device against the receiver band gap wavelength $\lambda_g$ is plotted in Fig. 2. With the BB emitter, a maximal efficiency of $\eta_{BB,max} \simeq 0.3$, known as Shockley-Queisser limit, is observed for $\lambda_g \simeq 1 \ \mu m$ (very close to crystalline silicon band gap wavelength). This upper limit is mainly due to two phenomena: the non conversion of photons with energy lower than $E_g$ and the thermal dissipation of the energy surplus of photons with energy higher than $E_g$. A low-pass filter would increase efficiency by eliminating low energy photons. The filter cut-off wavelength is $\lambda_{cutoff} = \lambda_{peak} = 1 \mu m$. A maximal efficiency $\eta_{filter,max} = 0.4$ can then be reached. To obtain this result, we make the assumption that radiation which is not transmitted by the filter is recovered by the system. Efficiency would be lower without this assumption. A SE goes far beyond by eliminating low energy photons and too energetic ones. The efficiency limit is increased up to $\eta_{SE,max} \simeq 0.6$ for $\lambda_g \simeq 1 \ \mu m$. In addition, the device configuration is simpler than the previous one. Besides, as it can be expected, a high and rapid efficiency drop is observed for the SE around $\lambda = 1.1 \ \mu m$. Moreover, since the emission is done over a very narrow wavelength band, efficiency drops faster than that of a system with a BB when the receiver band gap wavelength increases and moves away from the emission peak wavelength. A SE is then interesting only if it is coupled to a well adapted receiver.

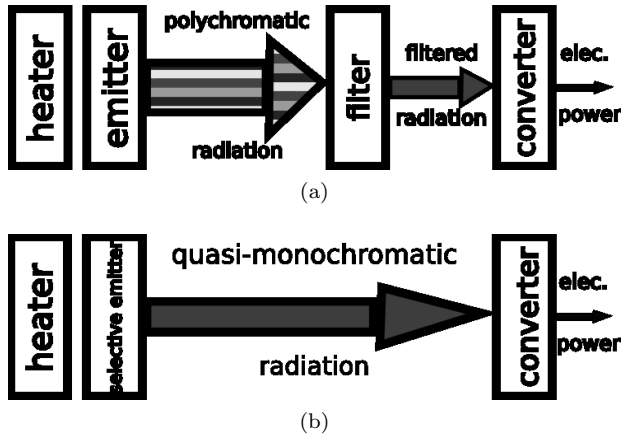

FIG. 1: A TPV device with a filtered blackbody emitter (a) and with a selective emitter (b)

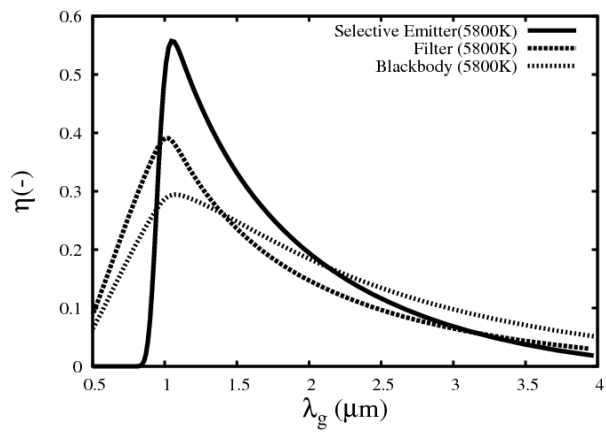

FIG. 2: Efficiency of a TPV device with a BB emitter, a filtered BB radiation and a SE versus the TPV cell band gap wavelength. Both of the BB and the SE are at 5800 K.

### 2. Effect of the emission peak width

A main difference between BB and SEs radiation at a fixed temperature, is that the latter can be tuned in different ways. To maximize the efficiency with the BB solar radiation, only one type of converters can be used, those with a band gap energy around $E_g = 1.1$ eV. Under these conditions, and with an incident radiation density $\Phi_i = 0.135$ W.cm$^{-2}$, the system maximal efficiency is $\eta_{BB,max} = 0.3$ and the maximal output power density $P_{BB,max} = 0.04$ W.cm$^{-2}$. By contrast however, a SE can be adapted to different converters by shifting the emission peak wavelength.

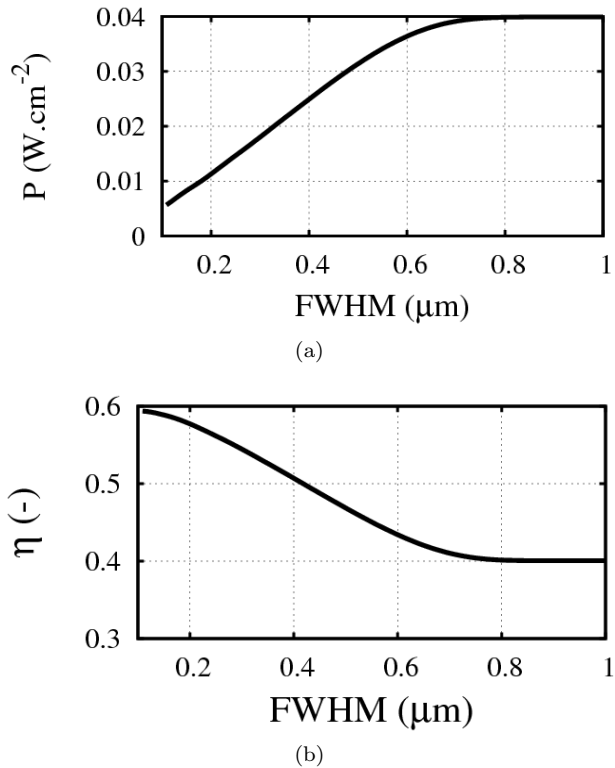

(a)

(b)

FIG. 3: A TPV device output power (a) and efficiency (b) versus the SE emission peak full width at half maximum (FWHM)

Besides, a large range of efficiencies and powers can be obtained if variable width emission peaks are considered. We calculate the efficiency $\eta$ and the output electric power $P_e$ of a TPV device with a selective emitter as given in equations 1 and 3 respectively function of the emission peak width. The full width at half maximum (FWHM) appears in the definition of $\epsilon_{SE}$ as follows:

$$\begin{cases} \epsilon_{SE}(\lambda) = 1 \text{ for } \lambda \in [1 - FWHM, 1]\mu m \\ \epsilon_{SE}(\lambda) = 0 \text{ otherwise} \end{cases} \tag{11}$$

FWHM is varied between 0.1 $\mu m$ and 0.9 $\mu m$. It mainly affects the calculation of the emitted radiation thus the values of radiation absorbed by the converter and the resulting electric power. We keep $E_g = 1.1$ eV. Fig. 3 presents the output power and efficiency of a TPV system against the emission peak FWHM. The same maximal power $P_{BB,max}$ can be reached with efficiency higher than $\eta_{BB,max} = 0.3$. Higher efficiencies can be reached, with lower output powers though. A compromise between power and efficiency has to be found for each application.

Use conditions of a SE and direct solar radiation are not the same. First, the view factor of a SE is much higher than this of the solar disk. Second, temperatures that can be reached with a SE are much lower than this of the solar surface temperature. Yet, this academic comparison is proposed for simplicity sake and in order to emphasize the potential usefulness of SEs. More realistic conditions will be considered later.

In the following paragraphs, an efficient optimization method for designing SEs with emission peaks at different wavelengths and with different widths is presented.

## III. SELECTIVE EMITTERS DESIGN AND OPTIMIZATION

### A. Problem to solve

In the previous paragraph, it has been reminded that a substantial part of the incident radiative energy is not converted into electricity with usual PV/TPV devices, in particular the part contained in photons with energy lower or too much higher than the converter band gap. In this paragraph, we propose a method to conceive and optimize emitters in order to prohibit the emission of such photons. In this work, only 1D multilayer structures are considered. A SE emitting in a narrow spectral band with a predefined FWHM around the peak maximum wavelength $\lambda_p$ is looked for. First, the emission peak position and its width are chosen arbitrary. We will show later that they can be tuned. The emissivity peak is modeled by a Gaussian function. The target optical properties are given by:

$$\begin{cases} \epsilon_{target}(\lambda, \theta) = \epsilon_{max} exp\left[\frac{(\lambda - \lambda_p)^2}{\alpha}\right] \text{ for } \theta \in [0, \frac{\pi}{2}] \\ \rho_{target}(\lambda, \theta) = 1 - \epsilon_{target}(\lambda, \theta) \end{cases} \quad (12)$$

where $\epsilon(\lambda, \theta)$ and $\rho(\lambda, \theta)$ figure the structure spectral and directional emissivity and reflectivity respectively. $\epsilon_{max} = 1$ is the maximal emissivity value and the parameter $\alpha$ allows the control of the Gaussian FWHM. The structure would obviously have a nil transmittance.

Since a structure with certain emissivity and reflectivity is looked for, the following objective function (Fitness) is to be minimized:

$$F = \sum_p \int_{\theta_1}^{\theta_2} \int_{\lambda_{min}}^{\lambda_{max}} \left[\epsilon_{target}(\lambda, \theta) - \epsilon_{struc}^p(\lambda, \theta)\right]^2 d\theta d\lambda$$

$$+ \sum_p \int_{\theta_1}^{\theta_2} \int_{\lambda_{min}}^{\lambda_{max}} \left[\rho_{target}(\lambda, \theta) - \rho_{struc}^p(\lambda, \theta)\right]^2 d\theta d\lambda$$

$$(13)$$

where the discrete sum operates over both thermal radiation polarization states. $\epsilon_{target}$ and $\rho_{target}$ are the desired emissivity and reflectivity and $\epsilon$ and $\rho$ those of the optimized structure. The structure that would exhibit such radiative properties is not known a priori. An inverse problem where the number of layers, their composition and arrangement are unknown is then solved. For a structure with $n$ layers, at least $2n$ parameters are to be determined: each layer composition and thickness. Stochastic optimization methods seem to be appropriate to deal with such high unknown number problems.

### B. Particle Swarm Optimization

Drevillon et al. successfully used genetic algorithms [6] to solve such problems but the obtained structures were quite complex (composed of a few dozens of layers). All the same, using a trial and error approach, some of these structures have been simplified and the number of layers reduced down to two. The feasibility of such simple 1D multilayer SEs was then proved [11]. In the following, a systematic method to find other solutions and to simplify them based on particle swarm optimization [10, 12] is presented.

Multilayer structures composed of semiconductors, metals and dielectrics have been considered. The number of layers and their properties (thicknesses, refractive indexes, dopant concentration, etc . . . ) are to be determined. For simplicity sake, two-material structures (that will be referred to as material **0** and material **1**) are considered. For example, an $n$-layer structure can be represented by a numerical sequence of the form $[t_1, \ldots, t_i, \ldots, t_n, d_1, \ldots, d_i, \ldots, d_n, \epsilon_0, \epsilon_1, c_0, c_1]$ where the $t_i \in [0..1]$ determines the layer $i$ type, $d_i$ determines its thickness, $\epsilon_0, \epsilon_1, c_0$ and $c_1$ the dielectric permittivities and the dopant concentrations (in the case of doped semiconductors) of materials **0** and **1** respectively.

A random population of $p$ particles is initially generated. Each particle (each particle designs a specific structure) has an initial position and velocity in the search space. In our case, it is a $2n + 4$-dimensions space. The coordinate according to each dimension defines the value of a parameter of the multilayer structure. Particles move in the search space in order to reach an optimum. At each step of their movement, the reflectance and emissivity of each particle/structure are calculated using transfer matrix formalism as previously detailed in [6]. These properties are then compared to the target structure properties according to equation 13. After comparison, the position $x_i$ and velocity $v_i$ of each particle $i$ are updated. The position update takes into account the velocity at the previous step. The velocity accounts for three components: a first one toward the best past position of the considered particle (is called memory or cognitive component), a second one toward the best past position of the particle neighborhood (is called social component) and finally the current velocity $v_i$ (inertial component). The best position is the position that minimizes $F$. The particle neighborhood is a subset of particles selected giving a predefined rule (it may be a subset of the nearest $s$ particles ($s \leq p$) to particle $i$ in the search space, it is then updated at each iteration, or it may be a fixed particles subset for the whole simulation). The new velocity and position are then given by:

$$v_{i+1} = \omega_1 v_i + \omega_2 r_1 (x_{bn,i} - x_i) + \omega_3 r_2 (x_{bm,i} - x_i) \quad (14)$$

$$x_{i+1} = x_i + v_{i+1} \quad (15)$$

where $\omega_1$, $\omega_2$ and $\omega_3$ are inertia, social and cognitive weight factors, $r_1$ and $r_2$ are two random factors. $x_{bn,i}$ and $x_{bm,i}$ are the best positions in the particle $i$ neighborhood and memory respectively. Weight factors were fixed to $\omega_1 = 0.729$, $\omega_2 = 1.494$, $\omega_3 = 1.494$ as recommended in literature [12] but can be varied if necessary. If physical parameters are looked for, all values are not allowed. When a particle runs out of the admissible search domain, it is put back on its boundary with a nil velocity. This operation is repeated until a satisfying solution or a maximal loop number is reached. Each iteration is equivalent to a unitary time step. This explains the apparent non homogeneity of the velocity and position quantities summed in equations 14 and 15.

## C. Results

The previous algorithm has been applied. Populations were always composed of 20 particles. 500 iterations were sufficient to obtain satisfying structures. During the optimization process, the maximal layers number, the lowest and highest layers thicknesses and the maximal dopant concentration were fixed to 50, 30 nm, 1000 nm and $5.10^{20}(\text{cm}^{-3})$ respectively. The materials optical properties are always considered at 300 K. Several structures with quasi-coherent emission at different wavelengths and with different emission peaks widths were obtained. Some bilayer and four-layer structures which are among the simplest ones, are presented.

### 1. Bilayer structures

Various bilayer structures exhibiting quasi-coherent emission have been obtained. The structures are made of a transparent material layer coating a lossy material thinner layer. The transparent material layer plays the role of an anti reflection coating [13] at the peak wavelength while the lossy material, if it is well chosen, can enhance the absorption thus the emission of the non reflected radiation around the anti-reflection wavelength. A bilayer anti reflection structure is presented in Fig. 4a. Let's consider this structure immersed in air. The refractive indexes of the air, the coating and the substrate are $n_{air} = 1$, $n_{coat}$ and $n_{sub}$ respectively. $d$ figures the thickness of the coating. Under certain conditions [11], This structure can exhibit nil reflectance at $\lambda = 4n_{coat}d$. The reflectance at normal incidence of such a structure, made of a 735 nm thick germanium (Ge) coating (germanium is transparent in mid infrared) on a substrate of a fictive transparent material which refractive index is equal to the real part of the refractive index of silicon carbide (SiC) is plotted in Fig. 5a. A reflectance minimum is observed at $\lambda \simeq 12.6$ $\mu$m. This wavelength of nil reflectance is tuned to correspond to the wavelength of resonance of the dielectric permittivity of SiC as shown in Fig. 5b. It can be shifted if necessary by varying the coating thickness. This structure exhibits neither absorption nor emission of light since both used materials are transparent.

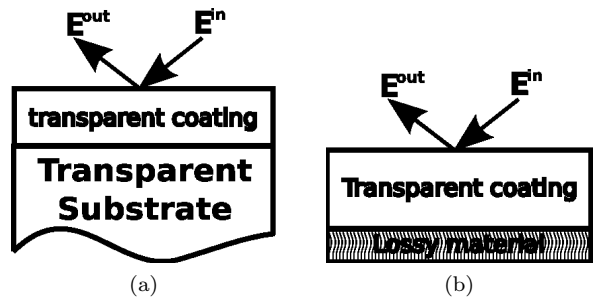

FIG. 4: (a) A bilayer antireflection structure. (b) A bilayer selective emitter.

The dielectric permittivity of SiC is given by the simple Lorentz oscillator[14] $\epsilon(\omega) = \epsilon_\infty \left[ 1 + \dfrac{(\omega_L^2 - \omega_T^2)}{\omega_T^2 - \omega^2 - i\Gamma\omega} \right]$ where $\omega_T = 14.937 \times 10^{13}\text{rad.s}^{-1}$, $\omega_L = 18.253 \times 10^{13}\text{rad.s}^{-1}$, $\Gamma = 8.966 \times 10^{11}$ and $\epsilon_\infty = 6.7$ are the transverse and the longitudinal optical phonon pulsation, the damping constant, and the high frequency dielectric constant, respectively. The substrate of the previous structure is replaced by a thin lossy SiC layer. This layer has to be thick enough to absorb all incident radiation around $\lambda_{peak} = 12.6$ $\mu$m. For this calculation, a 65 nm thick layer is considered. There is no important angular variation of radiative properties thus only quantities at normal incidence are presented here. In fact, all incident radiation between $0°$ and $90°$ is refracted within a very small $14°$ wide cone in the first layer since Ge refractive index ($n_{Ge} = 4$) presents a high contrast with air refractive index ($n_{air} = 1$). The reflectance and emissivity at normal incidence of the SE are plotted in Fig. 6.

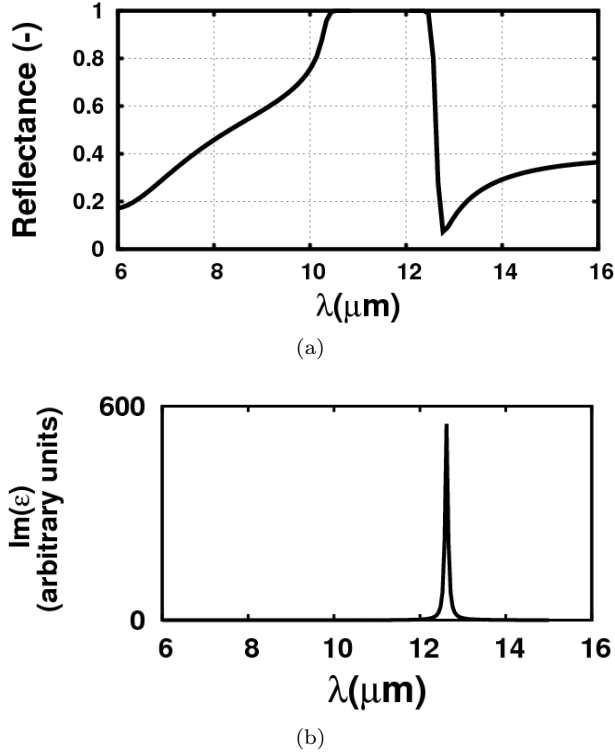

FIG. 5: (a) The reflectance of a thin germanium layer on a fictive transparent substrate. (b) The imaginary part of the dielectric permittivity of SiC, $\Im(\epsilon)$. The reflectance minimum wavelength corresponds to the dielectric permittivity peak wavelength.

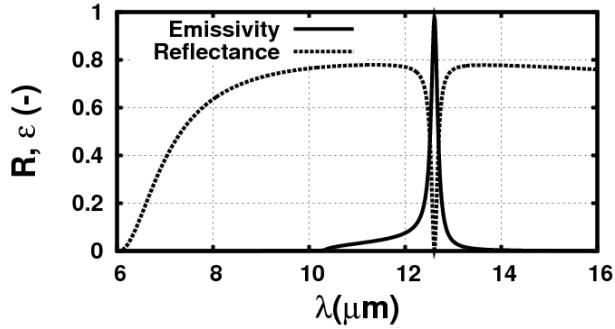

FIG. 6: Normal spectral reflectance and emissivity of a bilayer structure made of a 735 nm thick Ge layer and a 65 nm thick SiC layer.

Other structures based on the same physics have been obtained. The anti-reflection coating is always made of Germanium while Boron Nitride (BN) and Aluminium Nitride (AlN) are used as lossy materials. BN and AlN present strong lattice resonances in the mid infrared as well as SiC. The coating thickness is adjusted to make the reflectance minimum correspond to the dielectric permittivity resonance wavelength while the thickness of the lossy material is adjusted to ensure the absorption of all incident radiation around the emission peak wavelength. The emissivity at normal incidence of these structures is presented in Fig. 7. Their dimensions and radiative properties are summarized in table I. The dielectric permittivity of AlN is modeled by the simple Lorentz oscillator. The model parameters are $\omega_T = 12.346 \times 10^{13}$ rad.s$^{-1}$, $\omega_L = 17 \times 10^{13}$ rad.s$^{-1}$, $\Gamma = 1.88 \times 10^{12}$, $\epsilon_\infty = 4.77$. BN dielectric permittivity is modeled by a four parameters model $\epsilon(\omega) = \epsilon_\infty + 4\pi\rho\omega_0^2 \left[ \dfrac{1}{(\omega_0^2 - \omega^2) + i\Gamma\omega} \right]$ where $\epsilon_\infty = 4.5$, $4\pi\rho = 2.6$, $\omega_0 = 1065$ cm$^{-1}$ and $\Gamma = 40.5$ cm$^{-1}$ are the high frequency limiting value of the dielectric constant, the strength of the resonance, TO phonon-mode pulsation and the damping constant [14]. Using these different wide gap semiconductors, we are able to obtain quasi-coherent emission at different wavelengths and emission peaks with different widths. However, the possibilities with such structures are relatively limited since the emission peak wavelength and width are determined by the lossy material dielectric permittivity resonance that can hardly be controlled. Besides, the resonance of the dielectric permittivity that leads to the emission resonance is due to lattice vibrations resonances. Lattice vibrations generally correspond to far infrared wavelengths which are not interesting for TPV applications.

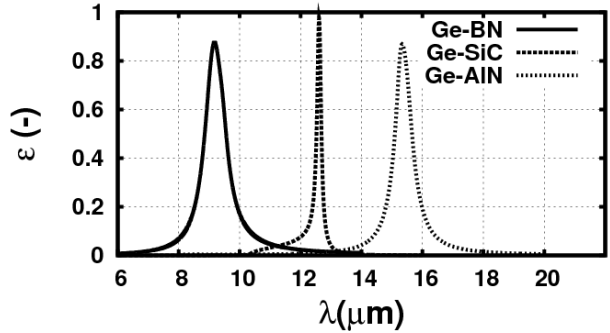

FIG. 7: Normal spectral emissivity of bilayer structures made of a germanium coating and a thin layer of BN, SiC and AlN.

| Coating | Lossy Mat. | $d_1$(nm) | $d_2$(nm) | $\lambda_p(\mu m)$ | FWHM($\mu m$) |
|---------|-----------|-----------|-----------|---------------------|----------------|
| Ge | SiC | 735 | 65 | 12.6 | 0.17 |
| Ge | BN | 450 | 310 | 9.2 | 0.9 |
| Ge | AlN | 950 | 300 | 15.3 | 0.7 |

Table I: Different bi-layer structures exhibiting quasi-coherent spectral emission in mid-IR. $d_1$ and $d_2$ are the coating and the lossy material thicknesses respectively. $\lambda_p$ and FWHM are the peak position and full width at half maximum.

### 2. Four-layer structures

Other structures, made of four layers, also exhibit coherent thermal radiation. These structures necessary alternate a lossless material and a lossy one. Fig. 8 shows a four-layer structure. Noble metals (gold and silver) and heavily doped semiconductors (Silicon) are used as lossy materials. Any transparent material in the considered wavelength range can play the role of the lossless material. Crystalline silicon is used in our simulations. Gold and silver dielectric

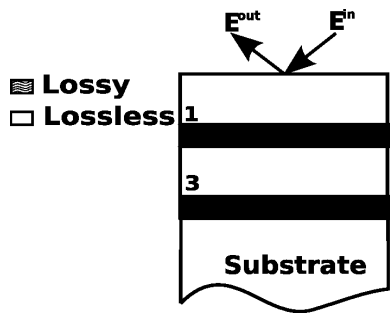

FIG. 8: A four-layer structure made by alternating a lossless and a lossy material layers.

permittivities are modeled by Drude models as can be found in usual reference literature [14]. Doped silicon dielectric permittivity is modeled by a Drude model [15] :

$$\epsilon(\omega) = \epsilon_\infty \left( 1 - \frac{\omega_p^2}{\omega(\omega + \frac{i}{\tau})} \right)$$ (16)

where $\omega = 2\pi c/\lambda$, $\omega_p^2 = \frac{Ne^2}{(m^*\epsilon_0\epsilon_\infty)}$ and $\epsilon_\infty$ are the angular frequency, the squared plasmon pulsation and the dielectric constant for high frequencies respectively. $\epsilon_\infty = 11.7$ for doped silicon. The carrier concentration $N$, the electron / hole effective mass $m^*$ and the relaxation time $\tau$ are doping-dependent. Their calculation details can be found in literature [15, 16]. Doping levels between $3.10^{19}$ cm$^{-3}$ and $5.10^{20}$ cm$^{-3}$ are considered. Crystalline silicon is lossless in the considered wavelength range with $\epsilon_{Si} = 11.7$ [14]. Structures with coherent emission in the near and mid-infrared are looked for. Target emissivity at normal incidence is given in Fig. 9a. It is the same for both light polarizations. Several solutions exhibiting such emissivity are obtained. The normal spectral emissivity of a Silicon/Gold structure (Si/Au/Si/Au) for both TE and TM polarizations is also presented in Fig. 9a. Curves corresponding to the two polarization states can hardly be distinguished since they coincide rather perfectly. The different layers thicknesses

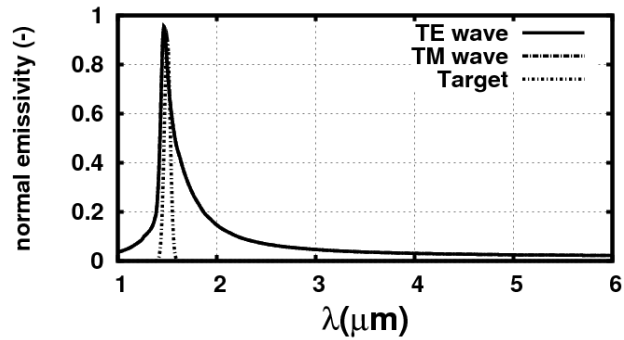

(a)

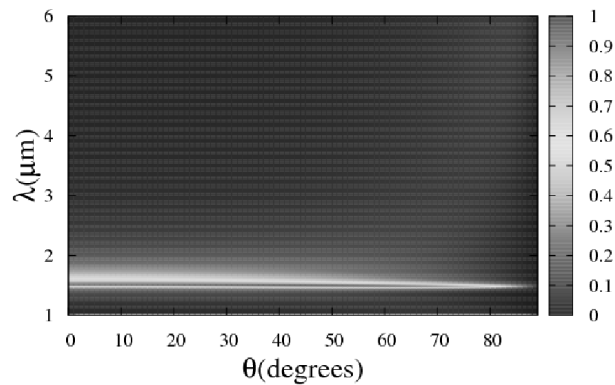

(b)

FIG. 9: (a) Target emissivity (equation 12) and spectral emissivities of the four-layer structure for both TE (Transverse Electric) and TM (Transverse Magnetic) light polarizations at normal incidence, obtained by Matrix Transfer method. (b) Spectral and directional emissivity of the four-layer structure for TM polarization.

are $d_1 = 90$ nm, $d_2 = 40$ nm, $d_3 = 170$ nm and $d_4 = 400$ nm. An emissivity peak is actually observed around 1.5 $\mu m$

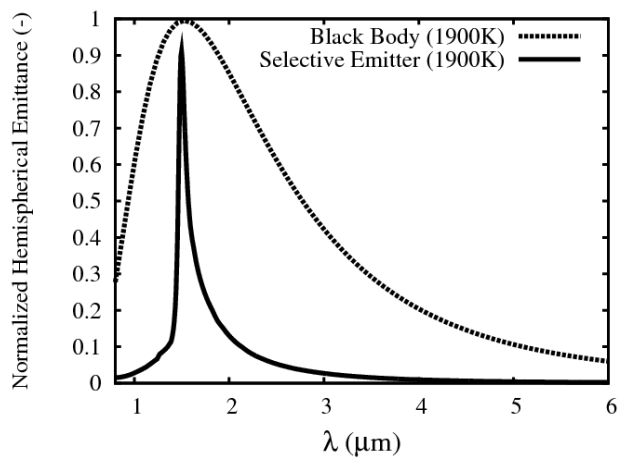

FIG. 10: Hemispherical emittance of a BB and the Gold/Silicon four-layer structure, both of them at 1900 K.

and nothing elsewhere as specified for the target. Fig. 9b presents the spectral and directional emissivity of the same structure for TM polarization. No significant angular dependance of the emissivity is observed. For comparison sake, we present in Fig. 10 the monochromatic hemispherical emittance (calculated according to equation 17) of this structure and that of a BB. Both of them are considered at $T = 1900$ K to make the emission peak wavelength coincide with the BB maximal emission wavelength given by Wien's displacement law. Although of comparable maximal emission intensities, the four-layer structure exhibits an emission peak much narrower than the BB spectrum. In fact, while 95% of the considered BB radiation is emitted between 0.75 $\mu$m and 6 $\mu$m, SE radiation is emitted between 1.44 $\mu$m and 1.56 $\mu$m if we consider its FWHM.

$$H_\lambda = \int_{\Omega=0}^{2\pi} \epsilon_{\lambda,\theta} \ I_{\lambda,\theta}(T_e) \ d\Omega \tag{17}$$

The emission peak of these structures is due to a Fabry-Perot cavity resonance [9] where the gold layers play the role of reflectors (noble metals are very good reflectors for infrared radiation) and the cavity is made of the lossless material. The resonance peaks wavelength at normal incidence are given by:

$$\lambda_m \simeq \frac{2n_c d_c}{m} \tag{18}$$

where $n_c$ and $d_c$ are the cavity refractive index and thickness respectively and $m$ is an integer that determine the resonance mode order. A rigorous equality does not appear in relation 18 because the peak wavelength also depends on the reflectors (Au layers) thicknesses and optical properties. The cavity optical thickness ($n_c d_c$ at normal incidence) allows the control of the peak wavelength. The lossy material choice allows to control the emission peak width. Doped silicon which dielectric permittivity can be tuned by the doping is considered in our simulations.

The BB temperature is chosen to make the first order resonance mode wavelength coincide with Wien's wavelength. Since the Fabry-Perot cavity resonance peaks are modulated by Planck's law, the first order peak is amplified and the higher order peaks that appear for shorter wavelengths are strongly attenuated.

Other structures that present emission peaks at other wavelengths and emission peaks of different widths are reported in tables II and III. Combining these data with (FWHM,$P$) and (FWHM,$\eta$) relations given in paragraph II B would allow to make the right choice for a given application. However, since proposed materials are not likely to bear temperatures as high as 5800 K, it is essential to consider lower and physically acceptable temperatures.

| n° | 0 | 1 | $d_1$(nm) | $d_2$(nm) | $d_3$(nm) | $d_4$(nm) | $c_1$(cm$^{-3}$) |
|---|---|---|---|---|---|---|---|
| 1 | Si | Doped Si | 375 | 48 | 194 | 30 | 3.72 |
| 2 | Si | Doped Si | 375 | 48 | 194 | 30 | 2.72 |
| 3 | Si | Doped Si | 375 | 48 | 194 | 30 | 1.72 |
| 4 | Si | Gold | 87 | 36 | 150 | 390 | – |
| 5 | Si | Gold | 87 | 36 | 200 | 390 | – |
| 6 | Si | Gold | 87 | 36 | 250 | 390 | – |

Table II: Different four-layer structures exhibiting coherent spectral emission in near IR obtained with the method explained above. Silicon/doped silicon and silicon/gold structures are presented here. The different layers thicknesses as well as the dopant concentration for doped silicon are determined.

| n° | $\lambda_p(\mu$m) | $FWHM(\mu$m) |
|---|---|---|
| 1 | 1.8 | 0.1 |
| 2 | 1.8 | 0.18 |
| 3 | 1.8 | 0.36 |
| 4 | 1.4 | 0.14 |
| 5 | 1.76 | 0.15 |
| 6 | 2.1 | 0.14 |

Table III: Characterization of emission peaks observed with structures presented in table II.

## IV. CONCLUSION

In this work, a method to conceive selective emitters for high efficiency TPV devices is proposed. Some simple bilayer and four-layer structures resulting from this optimization process and exhibiting coherent thermal emission at different wavelengths with variable width peaks are presented. Based on a simplified efficiency calculation, we showed that an appropriate TPV device using this kind of emitters clearly surpasses the same devices with a blackbody emitter. Besides, the proposed structures can be easily modified in order to be used with different receivers and to obtain different electrical powers. These results are admittedly encouraging but a more detailed efficiency calculation should be done taking into account other efficiency limiting phenomena (non ideal absorption, quantum efficiency, recombinations ...). In addition, used materials radiative properties were considered at 300 K while the emitter is supposed to be at much higher temperatures. The influence of temperature on both the emitter and the converter should be investigated. An experimental verification is currently being held.

## ACKNOWLEDGMENTS

We would like to acknowledge useful discussions with Philippe Ben-Abdallah, Jean-Jacques Greffet and Manuel Girault and gratefully acknowledge the support of the Agence Nationale de la Recherche through the Source-TPV project ANR 2010 BLAN 0928 01.